\documentclass[english]{article}
\usepackage[T1]{fontenc}
\usepackage[latin9]{inputenc}
\usepackage{geometry}
\usepackage{array}
\usepackage{color}
\usepackage{graphicx}

\makeatletter

\providecommand{\tabularnewline}{\\}

\makeatother

\usepackage{babel}

\begin{document}
\begin{center}
\textbf{\Large Generalized structure of higher order nonclassicality }
\par\end{center}{\Large \par}

\begin{center}
Amit Verma%
\footnote{amit.verma@jiit.ac.in%
} and Anirban Pathak%
\footnote{anirbanpathak@yahoo.co.in%
}
\par\end{center}

\begin{center}
Department of Physics and Materials Science and Engineering, JIIT
University, A-10, Sector-62, Noida, UP-201307, India
\par\end{center}

\begin{abstract}
A generalized notion of higher order nonclassicality (in terms of
higher order moments) is introduced. Under this generalized framework
of higher order nonclassicality, conditions of higher order squeezing
and higher order subpoissonian photon statistics are derived. A simpler
form of the Hong-Mandel higher order squeezing criterion is derived
under this framework by using an operator ordering theorem introduced
by us in {[}J. Phys. A. \textbf{33} (2000) 5607]. It is also generalized
for multi-photon Bose operators of Brandt and Greenberg. Similarly,
condition for higher order subpoissonian photon statistics is derived
by normal ordering of higher powers of number operator. Further, with
the help of simple density matrices, it is shown that the higher order
antibunching (HOA) and higher order subpoissonian photon statistics
(HOSPS) are not the manifestation of the same phenomenon and consequently
it is incorrect to use the condition of HOA as a test of HOSPS. It
is also shown that the HOA and HOSPS may exist even in absence of
the corresponding lower order phenomenon. Binomial state, \textcolor{black}{nonlinear
first order excited squeezed state (NLESS) and nonlinear vacuum squeezed
state (NLVSS) }are used as examples of quantum state and it is shown
that these states may show higher order nonclssical characteristics.
It is observed that the Binomial state which is always antibunched,
is not always higher order squeezed and NLVSS which shows higher order
squeezing does not show HOSPS and HOA. The opposite is observed in
NLESS and consequently it is established that the HOSPS and HOS are
two independent signatures of higher order nonclassicality.
\end{abstract}
~

Pacs: 42.50.-p, 42.50.Ar, 42.50.Lc


\section{Introduction: The generalized notion of higher order nonclassical
states}

A state which does not have any classical analogue is known as nonclassical
state. To be precise, when the Glauber Sudarshan P function of a radiation
field become negative or more singular than a delta function then
the radiation field is said to be nonclassical. In these situations
quasi probability distribution P is not accepted as classical probability
and thus we can not obtain an analogous classical state. For example,
squeezed state and antibunched state are well known nonclassical states.
These two lowest order nonclassical states have been studied since
long but the interest in higher order nonclassical states is relatively
new. \textcolor{black}{Possibilities of observing higher order nonclassicalities
in different physical systems have been investigated in recent past
{[}\ref{the:hong}-\ref{the:Amit}]. For example,} i) higher order
squeezed state of Hong Mandel type {[}\ref{the:hong}-\ref{the:gerry}],
ii) higher order squeezed state of Hillery type {[}\ref{hillery},\ref{the:Duc}],
iii) higher order subpoissonian photon state {[}\ref{the:Kim0}-\ref{the:Prakash-Mishra}]
and iv) higher order antibunched state \textcolor{black}{{[}\ref{lee1}-\ref{the:Amit}]
are recently studied in different physical systems. But the} general
nature of higher nonclassicality and the mutual relation between these
higher order nonclassical states have not been studied till now. Present
work aims to provide a general and simplified frame work for the study
of higher order nonclassical state\textcolor{black}{. }

\textcolor{black}{Commonly, second order moment (standard deviation)
of an observable is considered to be the most natural measure of quantum
fluctuation {[}\ref{the:Orlowski}] associated with that observable
and the reduction of quantum fluctuation below the coherent state
(poissonian state) level corresponds to lowest order nonclassical
state. For example, an electromagnetic field is said to be electrically
squeezed field if uncertainties in the quadrature phase observable
$X$ reduces below the coherent state level (i.e. $\left(\Delta X\right)^{2}<\frac{1}{2}$)
and antibunching is defined as a phenomenon in which the fluctuations
in photon number reduces below the Poisson level (i.e. $\left(\Delta N\right)^{2}<\langle N\rangle$)
{[}\ref{hbt},\ref{nonclassical}].} In Essenes, if we consider an
arbitrary quantum mechanical operator $A$ and a quantum mechanical
state $|\psi\rangle$ then the state $|\psi\rangle$ is lowest order
nonclassical with respect to the operator $A$ if \begin{equation}
(\Delta A)_{|\psi\rangle}^{2}<(\Delta A)_{|poissonian\rangle}^{2}.\label{eq:second-nonclassicality}\end{equation}
If $|\psi\rangle$ corresponds to an electromagnetic field, this condition
will mean that the radiation field is nonclassical. This condition
can now be further generalized and we can say that a state $|\psi\rangle$
has nth order nonclassicality with respect to the operator $A$ if
the $n$th order moment of $A$ in that state reduces below the value
of the $nth$ order moment of $A$ in a poissonian state, i.e. the
condition of $nth$ order nonclassicality is \begin{equation}
(\Delta A)_{|\psi\rangle}^{n}<(\Delta A)_{|poissonian\rangle}^{n},\label{eq:nth nonclassicality}\end{equation}
where $(\Delta A)^{n}$ is the $n$th order moment defined as \begin{equation}
\langle(\Delta A)^{n}\rangle=\sum_{r=0}^{n}\,^{n}C_{r}(-1)^{r}\overline{A^{r}}\,\,\bar{A}^{n-r}.\label{eq:nth moment}\end{equation}
If $A$ is a field operator then it can be expressed as a function
of creation and annihilation operators $a$ and $a^{\dagger}$ and
consequently further simplification of (\ref{eq:nth nonclassicality})
is possible by using the identity \begin{equation}
\langle:\left(A(a,a^{\dagger})\right)^{k}:\rangle_{|poisonian\rangle}=\langle\left(A(a,a^{\dagger})\right)\rangle_{|poisonian\rangle}^{k}\label{eq:simplification1}\end{equation}
where, the notation $:\,(A(a,a^{\dagger})^{k}:$ is simply a binomial
expansion in which powers of the $a^{\dagger}$ are always kept to
the left of the powers of the $a$. Here it would be interesting to
note that (\ref{eq:simplification1}) helps us to show that that the
Glauber Sudarshan P function is negative for condition (\ref{eq:nth nonclassicality}).
It is clear from (\ref{eq:nth moment}) that the problem of finding
out the $nth$ order moment of the operator $A$ essentially reduces
to a problem of operator ordering (normal ordering) of $A^{r}$. Here,
we would like to note that we observe the lowest order nonclassicality
for $n=2.$ And in this particular case ($n=2$) we obtain the condition
of squeezing of electric field, if $A=X=\frac{1}{\sqrt{2}}(a+a^{\dagger})$
and obtain the condition of antibunching if $A=N=a^{\dagger}a$. Now
if we need to generalize the idea of these well known lower order
nonclassical effects we have to find out normal ordered form of $X^{r}$
and $N^{r}$. In section 2 we start with an operator ordering theorem
which provides a normal ordered form of $X^{r}$ and obtain a simplified
expression of higher order squeezing%
\footnote{According to this notion of higher order squeezing Hillery type amplitude
powered squeezing is lower order squeezing of nonlinear bosonic operators
($Y_{1}$and $Y_{2}$). This is so because the amplitude powered squeezing
is described by the reduction (with respect to the poissonian state)
of second order moment of the corresponding quadrature variable. \textcolor{black}{One
can easily extend the existing notion of Hillery type squeezing and
obtain a new kind of higher order nonclassicality, namely, Hong Madel
type squeezing of Hillery type operator. }%
}. We have also generalized that expression for multi-photon Brandt-Greenberg
Bose operators. In the section 3 we have provided an operator ordered
form of $N^{r}$ and consequently obtained a condition for higher
order subpoissonian photon statistics. \textcolor{black}{In same section
we have also discussed the relation between different criteria of
higher order nonclassicalities. In section 4, }Binomial state, \textcolor{black}{nonlinear
first order excited squeezed state (NLESS) and nonlinear vacuum squeezed
state (NLVSS) }are used as examples of quantum state and it is shown
that these states may show higher order nonclssical characteristics.
\textcolor{black}{Finally section 5 is dedicated to conclusions.}

\section{Simplified condition for higher order squeezing:}

To obtain the condition for higher order squeezing (HOS) we need to
use the following operator order ordering theorem introduced by us
in {[}\ref{pathak-jphysa}]:\\
\textbf{Theorem 1:} If any two bosonic annihilation and creation operators
$a$ and $a^{\dagger}$ satisfy the commutation relation \begin{equation}
[a,a^{\dagger}]=1.\label{one}\end{equation}
Then for any integral values of $m$ \begin{equation}
(a^{\dagger}+a)_{N}^{m}=\sum_{r=0}^{\frac{m}{2}}t_{2r}\,^{m}C_{2r}:\,(a^{\dagger}+a)^{m-2r}:\label{th2.1}\end{equation}
with\begin{equation}
t_{2r}=\frac{2r!}{2^{r}(r)!}=(2r-1)!!=2^{r}\left(\frac{1}{2}\right)_{r}\label{th2.2}\end{equation}
where, the subscript $N$ stands for the normal ordering %
\footnote{One can write $f(a,a^{\dagger})$ in such a way that all powers of
$a^{\dagger}$ always appear to the left of all powers of $a$. Then
$f(a,a^{\dagger})$ is said to be normal ordered. %
}, $(x)_{r}$ is conventional Pochhammer symbol and the double factorial
is defined as \begin{equation}
n!!=\left\{ \begin{array}{c}
n(n-2)..5.3.1\, for\, n>0\, odd\\
n(n-2)..6.4.2\, for\, n>0\, even\\
1\, for\, n=-1,0\end{array}\right..\label{eq:double factorial}\end{equation}
This theorem of normal ordering is not restricted to the ordering
of annihilation and creation operator rather it is valid for any arbitrary
operator $E^{+}$ and its conjugate $E^{-}$ which satisfy, \begin{equation}
[E^{+},E^{-}]=C.\label{eq:one.1}\end{equation}
This is so because (\ref{eq:one.1}) can easily be reduced to the
form of (\ref{one}) as $[\frac{E^{+}}{\sqrt{C}},\frac{E^{-}}{\sqrt{C}}]=1$.
Apart from theorem 1 we need the following identity to proceed further:

\textcolor{black}{\begin{equation}
\begin{array}{ccc}
\,^{n}C_{r}\,^{r}C_{j} & = & \frac{n!}{(n-r)!r!}\frac{r!}{(r-j)!j!}=\frac{n!}{(n-j)!j!}\frac{(n-j)!}{[(n-j)-(r-j)]!(r-j)!}=\,^{n}C_{j}\,^{n-j}C_{r-j}.\end{array}\label{eq:identity1}\end{equation}
}Using (\ref{eq:nth moment}), (\ref{th2.1}), (\ref{th2.2}) and
(\ref{eq:identity1}) the $nth$ order moment of $\Delta E=E-\bar{E}$
(where the quadrature variable $E=(a+a^{\dagger})$) can be expressed
as \begin{equation}
\begin{array}{lcc}
\langle(\Delta E)^{n}\rangle & = & \sum_{r=0}^{n}\,^{n}C_{r}(-1)^{r}\overline{E^{r}}\,\,\bar{E}^{n-r}\\
 & = & \sum_{r=0}^{n}\,^{n}C_{r}(-1)^{r}\langle(a^{\dagger}+a)^{r}\rangle\langle a^{\dagger}+a\rangle^{n-r}\\
 & = & \sum_{r=0}^{n}\,^{n}C_{r}(-1)^{r}\sum_{i=0}^{\frac{r}{2}}t_{2i}\,^{r}C_{2i}\langle:\,(a^{\dagger}+a)^{r-2i}:\rangle\langle a^{\dagger}+a\rangle^{n-r}\\
 & = & \sum_{r=0}^{n}\sum_{i=0}^{\frac{r}{2}}t_{2i}\,^{n}C_{2i}\,^{n-2i}C_{r-2i}(-1)^{r-2i}\langle:\,(a^{\dagger}+a)^{r-2i}:\rangle\langle a^{\dagger}+a\rangle^{\{(n-2i)-(r-2i)\}}\\
 & = & \sum_{i=0}^{\frac{n}{2}}t_{2i}\,^{n}C_{2i}\langle:(\Delta E)^{n-2i}:\rangle.\\
\\\end{array}\label{eq:hong-mandel, proof}\end{equation}
Now if we follow (\ref{eq:nth nonclassicality}), and define nth order
squeezed state as a quantum mechanical state in which nth order moment
$\langle(\Delta E)^{n}\rangle$ is shorter than its poissonian sate
value then the condition for nth order squeezing reduces to \begin{equation}
\langle(\Delta E)^{n}\rangle<t_{n}=(n-1)!!\label{eq:cond1}\end{equation}
which can be alternatively written as \begin{equation}
\sum_{i=0}^{\frac{n}{2}-1}t_{2i}\,^{n}C_{2i}\langle:(\Delta E)^{n-2i}:\rangle<0.\label{eq:cond2}\end{equation}
or, \begin{equation}
\langle(\Delta E)^{n}\rangle=\sum_{r=0}^{n}\sum_{i=0}^{\frac{r}{2}}\sum_{k=0}^{r-2i}(-1)^{r}t_{2i}\,^{r-2i}C_{k}\,^{n}C_{r}\,^{r}C_{2i}\langle a^{\dagger}+a\rangle^{n-r}\langle a^{\dagger k}a^{r-2i-k}\rangle<(n-1)!!.\label{eq:cond3}\end{equation}
Conditions (\ref{eq:cond1}) and (\ref{eq:cond2}) coincide exactly
with the definition of Hong Mandel squeezing, reported in earlier
works %
\footnote{If we choose $E_{1}=E^{+}+E^{-}$ in analogy with Hong and Mandel
{[}\ref{the:hong}] then (\ref{eq:cond1}) reduces to \[
\langle(\Delta E)_{1}^{n}\rangle<t_{n}C^{\frac{n}{2}}=(n-1)!!C^{\frac{n}{2}},\]
where $n$ is even. This is the generalized expression obtained in
{[}\ref{the:hong}] by using some other trick. %
} {[}\ref{the:hong}, \ref{the:gerry}] and the equivalent condition
(\ref{eq:cond3}) considerably simplifies the calculation of HOS.
Now instead of $E$ if we calculate the $nth$ order moment for usual
quadrature variable $X$ defined as $X=\frac{1}{\sqrt{2}}(a+a^{\dagger})$,
then we obtain \begin{equation}
\langle(\Delta X)^{n}\rangle<\frac{1}{2^{\frac{n}{2}}}t_{n}=\frac{1}{2^{\frac{n}{2}}}(n-1)!!=\left(\frac{1}{2}\right)_{\frac{n}{2}}\label{eq:cond1.1}\end{equation}
or, \begin{equation}
\sum_{r=0}^{n}\sum_{i=0}^{\frac{r}{2}}\sum_{k=0}^{r-2i}(-1)^{r}\frac{1}{2^{\frac{n}{2}}}t_{2i}\,^{r-2i}C_{k}\,^{n}C_{r}\,^{r}C_{2i}\langle a^{\dagger}+a\rangle^{n-r}\langle a^{\dagger k}a^{r-2i-k}\rangle<\left(\frac{1}{2}\right)_{\frac{n}{2}}.\label{eq:cond2.1}\end{equation}
Starting from the generalized notion of higher order nonclassicality
(\ref{eq:nth nonclassicality}) we have obtained a closed from expression
of Hong-Mandel squeezing with the help of Theorem 1. The use of Theorem
1 not only simplifies the condition it significantly reduces the calculational
difficulties. To be precise, to study the possibility of HOS for an
arbitrary quantum state $|\psi\rangle$ we just need to calculate
$\langle a^{\dagger}+a\rangle$ and $\langle a^{\dagger k}a^{r-2i-k}\rangle$.
Calculation of this expectation values are simple. For example, if
we can expand the arbitrary state $|\psi\rangle$ in the number state
basis as \begin{equation}
|\psi\rangle=\sum_{j=0}^{N}C_{j}|j\rangle.\label{eq:gen1}\end{equation}
Then we can easily obtain, \begin{equation}
\langle\psi|a^{\dagger k}a^{r-2i-k}|\psi\rangle=\sum_{j=0}^{N-Max[k,\, r-2i-k]}C_{j+k}^{*}C_{j+r-2i-k}\frac{1}{j!}\left((j+k+r-2i)!(j+k)!\right)^{\frac{1}{2}}\label{eq:gen2}\end{equation}
where Max yields the largest element from the list in its argument
and \begin{equation}
\langle a^{\dagger}+a\rangle=\sum_{m=0}^{N-1}\sqrt{(m+1}\left(C_{m}C_{m+1}^{*}+C_{m}^{*}C_{m+1}\right).\label{eq:gen3}\end{equation}
Therefore, \begin{equation}
\begin{array}{lcl}
\langle(\Delta X)^{n}\rangle & = & \sum_{r=0}^{n}\sum_{i=0}^{\frac{r}{2}}\sum_{k=0}^{r-2i}(-1)^{r}\frac{1}{2^{\frac{n}{2}}}t_{2i}\,^{r-2i}C_{k}\,^{n}C_{r}\,^{r}C_{2i}\\
 & \times & \left(\sum_{m=0}^{N-1}\sqrt{(m+1}\left(C_{m}C_{m+1}^{*}+C_{m}^{*}C_{m+1}\right)\right)^{n-r}\\
 & \times & \sum_{j=0}^{N-Max[k,\, r-2i-k]}C_{j+k}^{*}C_{j+r-2i-k}\frac{1}{j!}\left((j+k+r-2i)!(j+k)!\right)^{\frac{1}{2}}.\end{array}\label{eq:gen4}\end{equation}
 In general, if we know the effect of $a^{s}$ on the state $|\Psi\rangle$
and the orthogonality conditions $\langle\Psi^{\prime}|\Psi\rangle$
then we can easily find out $\langle(\Delta X)^{n}\rangle$. Further,
since (\ref{eq:gen4}) is a $C$-number equation, analytical tools
like MAPPLE and MATHEMATICA can also be used to study the possibilities
of observing higher order squeezing (or higher order nonclassicality
in general). \textcolor{black}{This point will be more clear in section
4, where we will provide specific examples. Here we would like to
note that we can normalize (\ref{eq:cond1.1}) and rewrite the condition
of HOS as \begin{equation}
S_{HM}(n)=\frac{\langle(\Delta X)^{n}-\left(\frac{1}{2}\right)_{\frac{n}{2}}}{\left(\frac{1}{2}\right)_{\frac{n}{2}}}<0\label{eq:cond:new}\end{equation}
 where the subscript $HM$ stand for Hong Mandel.}

\subsection{Brandt-Greenberg operators and k-photon coherent state:}

The k-photon coherent state was introduced by D'Arino and coworkers
by using Brandt-Greenberg multi-photon operators {[}\ref{the:Ariano}]
$A_{k}$ and $A_{k}^{\dagger}$, which are defined as \begin{equation}
A_{k}^{\dagger}=\left[\left[\frac{N}{k}\right]\frac{N-k}{N}\right]a^{\dagger k},\label{eq:brandt-greenberg1}\end{equation}
\begin{equation}
A_{k}=(A_{k}^{\dagger})^{\dagger},\label{eq:bg2}\end{equation}
where the function {[}x] is defined as the greatest integer less or
equal to x; $a^{\dagger}$and $a$ are the usual bosonic relation
and annihilation operator and $N=a^{\dagger}a$ is the number operator.
This particular from of Brandt-Greenberg operators is also used in
the work of Buzek and Jex {[}\ref{the:buzek and jex}] in which they
have studied the amplitude $k-th$ power squeezing of the k-photon
coherent states. These operators satisfy the commutation relation
analogous to (\ref{one}), i.e. they satisfy, \begin{equation}
[A_{k},A_{k}^{\dagger}]=1.\label{eq:commutation2}\end{equation}
If any operator and its hermitian conjugate satisfies this kind of
commutation relation then it has to satisfy the operator ordering
theorem 1 and consequently we will be able to define Hong-Mandel squeezing
in terms that particular operator (in a modified Fock space). For
example, if we define to quadrature variables $X_{1k}$ and $X_{2K}$
as \begin{equation}
\begin{array}{lcl}
X_{1k} & = & A_{k}+A_{k}^{\dagger}\\
X_{2k} & = & A_{k}-A_{k}^{\dagger}\end{array}\label{eq:quadrature1}\end{equation}
 then we can define the condition for $nth$ order Hong-Mandel squeezing
as \begin{equation}
\sum_{r=0}^{n}\sum_{i=0}^{\frac{r}{2}}\sum_{k=0}^{r-2i}(-1)^{r}\frac{1}{2^{\frac{n}{2}}}t_{2i}\,^{r-2i}C_{k}\,^{n}C_{r}\,^{r}C_{2i}\langle A^{\dagger}+A\rangle^{n-r}\langle A^{\dagger k}A^{r-2i-k}\rangle<\left(\frac{1}{2}\right)_{\frac{n}{2}}.\label{eq:h1}\end{equation}
 This provides an extended notion of Hong-Mandel squeezing in a modified
Hilbert space.

\section{Higher order subpoissonian photon statistics}

In analogy to the procedure followed to derive the Hong-Mandel higher
order squeezing condition from the generalized expression (\ref{eq:nth nonclassicality})
of higher order nonclassicality, we wish to study the nonclassicality
associated with $A(a,a^{\dagger})=N=a^{\dagger}a.$ As we have already
discussed, for this purpose we will require operator ordered form
of $N^{r}$. Since the operator ordered expansion of $N^{r}$ will
not contain any off-diagonal term so it is justified to assume that
the normal ordered form of $(N)^{r}$can be given as \begin{equation}
N^{r}=\sum_{i=1}^{r}C_{r,i}:N^{i}:=\sum_{i=1}^{r}C_{r,i}a^{\dagger i}a^{i}.\label{eq:N^r-normal}\end{equation}
From this equation it is clear that $C_{r,1}=C_{r,r}=1$ and we can
write $N^{r+1}$as \begin{equation}
N^{r+1}=\sum_{i=1}^{r+1}C_{r+1,i}a^{\dagger i}a^{i}=\sum_{i=1}^{r}C_{r,i}a^{\dagger i}a^{i}a^{\dagger}a=\sum_{i=1}^{r}\left(C_{r,i}a^{\dagger i+1}a^{i+1}+iC_{r,i}a^{\dagger i}a^{i}\right)\label{eq:N^r+1}\end{equation}
 where the operator ordering identity, $a^{l}a^{\dagger}=a^{\dagger}a^{l}+la^{l-1}$
is used. Now, we will be able to obtain closed form normal ordered
expansion of $N^{r}$ provided we know the solution of the recurrence
relation: \begin{equation}
C_{r+1,i}=iC_{r,i}+C_{r,i-1}\label{eq:recurance1}\end{equation}
with $C_{r,0}=0$ and $C_{r,1}=1$. One can easily identify (\ref{eq:recurance1})
as the famous recurrence relation of Stirling number of second kind
{[}\ref{the:table of integrals}]. Thus we can write \begin{equation}
N^{r}=\sum_{k=1}^{r}S_{2}(r,k)a^{\dagger k}a^{k}=\sum_{k=1}^{r}S_{2}(r,k):N^{k}:=\sum_{k=1}^{r}S_{2}(r,k)N^{(k)},\label{eq:operaotr-orderredN^r}\end{equation}
where $S_{2}(r,k)$ is the Stirling number of second kind $N^{(k)}=a^{\dagger k}a^{k}$
is the $kth$ factorial moment of the number operator $N$. Now using
(\ref{eq:nth nonclassicality}), (\ref{eq:nth moment}), (\ref{the:Duc})
and (\ref{eq:operaotr-orderredN^r}) we can obtain the condition of
higher order subpoissonian photon statistics as \[
\langle(\Delta N)^{n}\rangle=\sum_{r=0}^{n}\,^{n}C_{r}(-1)^{r}\bar{N^{r}}\bar{N}^{n-r}=\sum_{r=0}^{n}\sum_{k=1}^{r}S_{2}(r,k)\,^{n}C_{r}(-1)^{r}\langle N^{(k)}\rangle\langle N\rangle^{n-r}<\langle(\Delta N)^{n}\rangle_{|poissonain\rangle}\]
or, \begin{equation}
d_{h}(n-1)=\sum_{r=0}^{n}\sum_{k=1}^{r}S_{2}(r,k)\,^{n}C_{r}(-1)^{r}\langle N^{(k)}\rangle\langle N\rangle^{n-r}-\sum_{r=0}^{n}\sum_{k=1}^{r}S_{2}(r,k)\,^{n}C_{r}(-1)^{r}\langle N\rangle^{k+n-r}<0.\label{eq:cond-HOSPS}\end{equation}
The negativity of $d_{h}(n-1)$ will mean $(n-1)th$ order subpoissonian
photon statistics. This condition is equivalent to the condition of
HOSPS obtained Mishra-Prakash {[}\ref{the:Prakash-Mishra}].

\subsection{Relation between the criteria of HOA and HOSPS}

\textcolor{black}{The criterion of HOA is expressed in terms of higher
order factorial moments of number operator. There exist several criterion
for the same which are essentially equivalent. Here we would like
to investigate how are they related to the criterion of HOSPS. Initially,
using the negativity of P function and theory of Majorization, Lee
{[}\ref{lee1}, \ref{lee2}] introduced the criterion for HOA as }

\textcolor{black}{\begin{equation}
R(l,m)=\frac{\left\langle N_{x}^{(l+1)}\right\rangle \left\langle N_{x}^{(m-1)}\right\rangle }{\left\langle N_{x}^{(l)}\right\rangle \left\langle N_{x}^{(m)}\right\rangle }-1<0,\label{eq:ho3}\end{equation}
where $N$ is the usual number operator, $\left\langle N^{(i)}\right\rangle =\left\langle N(N-1)...(N-i+1)\right\rangle $
is the $ith$ factorial moment of number operator, $l$ and $m$ are
integers satisfying the conditions $1\leq m\leq l$ and the subscript
$x$ denotes a particular mode. Ba An {[}\ref{ba an}] choose $m=1$
and reduced the criterion of $l$th order antibunching to \begin{equation}
A_{x,l}=\frac{\left\langle N_{x}^{(l+1)}\right\rangle }{\left\langle N_{x}^{(l)}\right\rangle \left\langle N_{x}\right\rangle }-1<0\label{eq:bhuta1}\end{equation}
or, \begin{equation}
\left\langle N_{x}^{(l+1)}\right\rangle <\left\langle N_{x}^{(l)}\right\rangle \left\langle N_{x}\right\rangle .\label{eq:ba an (cond)}\end{equation}
We can further simplify (\ref{eq:ba an (cond)}) as \begin{equation}
\left\langle N_{x}^{(l+1)}\right\rangle <\left\langle N_{x}^{(l)}\right\rangle \left\langle N_{x}\right\rangle <\left\langle N_{x}^{(l-1)}\right\rangle \left\langle N_{x}\right\rangle ^{2}<\left\langle N_{x}^{(l-2)}\right\rangle \left\langle N_{x}\right\rangle ^{3}<...<\left\langle N_{x}\right\rangle ^{l+1}\label{eq:ineq}\end{equation}
and obtain the condition for $l-th$ order antibunching as \begin{equation}
d(l)=\left\langle N_{x}^{(l+1)}\right\rangle -\left\langle N_{x}\right\rangle ^{l+1}<0.\label{eq:ho21}\end{equation}
This simplified criterion (\ref{eq:ho21}) coincides exactly with
the physical criterion of HOA introduced by Pathak and Garica {[}\ref{the:martin}]
and the criterion of Erenso, Vyas and Singh {[}\ref{singh1}]. In
{[}\ref{the:martin}] it is already shown that the depth of nonclassicality
of an $lth$ order antibunching is always more than that of $(l-1)th$
order antibunching of the same state. Consequently, \begin{equation}
d(l)<d(l-1)\label{eq:aa}\end{equation}
or, \begin{equation}
\langle N^{(l+1)}\rangle\langle N\rangle^{n-r}-\langle N\rangle^{l+1+n-r}<\langle N^{(l)}\rangle\langle N\rangle^{n-r}-\langle N\rangle^{l+n-r}.\label{eq:equivalence 1}\end{equation}
Now the condition for HOSPS, i.e. (\ref{eq:cond-HOSPS}) can be written
as} \begin{equation}
d_{h}(n-1)=\sum_{r=0}^{n}\sum_{k=1}^{r}S_{2}(r,k)\,^{n}C_{r}(-1)^{r}\left[\langle N^{(k)}\rangle-\langle N\rangle^{k}\right]\langle N\rangle^{n-r}=\sum_{r=0}^{n}\sum_{k=1}^{r}S_{2}(r,k)\,^{n}C_{r}(-1)^{r}d(k-1)\langle N\rangle^{n-r}<0.\label{eq:cond-HOSPS2}\end{equation}
\textcolor{black}{Above relation connects the condition of HOA (\ref{eq:ho21})
and that of HOSPS (\ref{eq:cond-HOSPS}) but does not provide any
conclusion about the mutual satisfiability. Physically it is expected
from the analogy of lower order phenomenon that all states that show
HOA should show HOSPS but the reverse should not be true. We have
not succeed in showing that analytically but we can establish that
with the help of simple density matrix of the form $\frac{1}{2}(|a\rangle\langle a|+|b\rangle\langle b|)$,
where $|a\rangle$and $|b\rangle$ are Fock states. The results are
shown in the Table 1. }

\textcolor{black}{}%
\begin{table}
\begin{tabular}{|c|c|c|c|c|>{\centering}p{2in}|}
\hline
Density matrix & Antibunching & SPS & HOA $(l=3)$ & HOSPS $(n=4)$ & Conclusions\tabularnewline
\hline
\hline
$\frac{1}{2}(|3\rangle\langle3|+|8\rangle\langle8|)$ & No & No & Yes & Yes & HOA and HOSPS can exist in absence of lower order\tabularnewline
\hline
$\frac{1}{2}(|4\rangle\langle4|+|10\rangle\langle10|)$ & No & No & No & Yes & HOA and HOSPS are different phenomenon\tabularnewline
\hline
\end{tabular}

\textcolor{black}{\caption{HOA and HOSPS are not the manifestation of the same phenomenon and
consequently it is incorrect to use the condition of HOA as a test
of HOSPS.}
}
\end{table}
\textcolor{black}{All the criterion related to HOA and HOSPS essentially
lead to same kind of nonclassicality which belong to the class of
strong nonclassicality according to the classification scheme of Arvind}
\textcolor{black}{\emph{et al}} \textcolor{black}{{[}\ref{the:Arvind}].
The Table 1 shows that HOA and HOSPS may be present in a system in
absence of corresponding lower order phenomenon. It also shows that
HOA and HOSPS are not the same phenomenon. To be precise, HOSPS can
be present in a system even in absence of HOA. Thus it is not proper
to consider the condition on HOA as the condition of HOSPS. In {[}}\ref{the:Duc}]
Duc has recently used criterion of HOA to study possibilities of observing
HOSPS in photon added coherent state. Incorrect choice of criterion
may yield incorrect conclusions so we need to be very careful before
choosing a criterion of higher order nonclassicality. Further in section
4 we have shown that the binomial state is not always higher order
squeezed but is always higher order antibunched. Thus we can conclude
that although they may be derived from same generalized framework
they are essentially independent criterion. This is in the sense that
\textcolor{black}{fulfillment} of one does not mean \textcolor{black}{fulfillment}
of the other.

\section{Examples}

\subsection{Binomial State}

An intermediate state is a quantum state which reduces to two or more
distinguishably different states (normally, distinguishable in terms
of photon number distribution) in different limits. In 1985, such
a state was first time introduced by Stoler \emph{et al.} {[}\ref{the:D-Stoler,-B}].
To be precise, they introduced Binomial state (BS) as a state which
is intermediate between the most nonclassical number state $|n'\rangle$
and the most classical coherent state $|\alpha\rangle$. They defined
BS as \begin{equation}
\begin{array}{lr}
|p,M\rangle=\sum_{n'=0}^{M} & B_{n'}^{M}\end{array}|n'\rangle=\sum_{n'=0}^{M}\sqrt{^{M}C_{n'}p^{n'}(1-p)^{M-n'}}|n'\rangle\,\,\,\:0\leq p\leq1.\label{eq:binomial1}\end{equation}
This state%
\footnote{The state is named as binomial state because the photon number distribution
associated with this state $\left(i.e.\,|B_{n}^{M}|^{2}\right)$is
simply a binomial distribution.%
} is called intermediate state as it reduces to number state in the
limit $p\rightarrow0$ and $p\rightarrow1$ (as $|0,M\rangle=0$ and
$|1,M\rangle=|M\rangle$) and in the limit of $M\rightarrow\infty,\, p\rightarrow1$,
where $\alpha$ is a real constant, it reduces to a coherent state
with real amplitude. Since the introduction of BS as an intermediate
state it was always been of interest to quantum optics, nonlinear
optics, atomic physics and molecular physics community. Consequently,
different properties of binomial states have been studied {[}\ref{the:VIdiella-Barraco}-\ref{the:OEBS}].
In these studies it has been observed that the nonclassical phenomena
(such as, antibunching, squeezing and higher order squeezing) can
be seen in BS.

Using the above definition of BS we obtain\begin{equation}
d(l)_{BS}=\frac{M!p^{l+1}}{(M-l-1)!}-(Mp)^{l+1},\label{eq:ami1}\end{equation}
\begin{equation}
d_{h}(n-1)_{BS}=\sum_{r=0}^{n}\sum_{k=1}^{r}\left[S_{2}(r,k)\,^{n}C_{r}(-1)^{r}(Mp)^{n-r}\left(\frac{M!p^{k}}{(M-k)!}-(Mp)^{k}\right)\right],\label{eq:ami2}\end{equation}
and \begin{equation}
\begin{array}{lcl}
S_{HM}(n)_{BS} & = & \frac{1}{\left(\frac{1}{2}\right)_{\frac{n}{2}}}\sum_{r=0}^{n}\sum_{i=0}^{\frac{r}{2}}\sum_{k=0}^{r-2i}(-1)^{r}\frac{1}{2^{\frac{n}{2}}}t_{2i}\,^{r-2i}C_{k}\,^{n}C_{r}\,^{r}C_{2i}\langle a^{\dagger}+a\rangle^{n-r}\langle a^{\dagger k}a^{r-2i-k}\rangle-1\\
 & = & \frac{1}{\left(\frac{1}{2}\right)_{\frac{n}{2}}}\left[\sum_{r=0}^{n}\sum_{i=0}^{\frac{r}{2}}\sum_{k=0}^{r-2i}(-1)^{r}\frac{1}{2^{\frac{n}{2}}}t_{2i}\,^{r-2i}C_{k}\,^{n}C_{r}\,^{r}C_{2i}\:\left[2(Mp)^{1/2}\sum_{n'=0}^{^{M-1}}B_{n'}^{M}B_{n'}^{M-1}\right]^{^{n-r}}\right.\\
 &  & \left.\left[\frac{M!^{2}p^{r-2i}}{(M-k)!(M-r+2i+k)!}\right]^{1/2}\sum_{n'=0}^{^{M-Max[k,r-2i-k]}}B_{n'}^{M-k}B_{n'}^{M-r+2i+k}\right]-1\end{array}\label{eq:ami3}\end{equation}

Above expressions are graphically represented in Fig. 1-Fig. 2. From
the Fig. 1 it is clear that the BS shows HOA and HOSPS simultaneously,
but they are not proportional to each other. This result is not of
much interest as we have already shown in {[}\ref{the:Amit}] that
the BS is always higher order antibunched and as every higher order
nonclassical state is expected to show HOSPS, independent of whether
they show HOA and Hong-Mandel squeezing or not. The result of real
physical relevance appears when we look at Fig. 2 which shows that
the BS does not show Hong Mandel squeezing for all values of $p$.
For example 4th order Hong Mandel squeezing vanishes for $M=50$ and
$p\geq0.8607$. In this range the state is still nonclassical and
shows HOA and HOSPS but does not show Hong Madel Squeezing. Consequently
we can conclude that the HOA and Hong Mandel squeezing are two independent
processes which may or may not appear together. Further it can be
observed that for the same photon member ($M$), the region of nonclassicality
decreases with the increase in order of Hong Mandel squeezing. To
be precise, when $M=50$ then $S_{HM}(4)_{BS}$ is negative till $p=0.8607,$
but $S_{HM}(6)_{BS}$ is negative till $p=0.7943$ and $S_{HM}(8)_{BS}$
is negative till $p=0.7343$.

\begin{figure}

\begin{centering}
\includegraphics{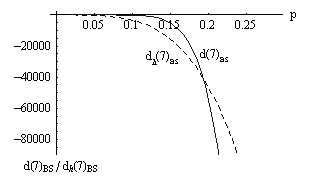}
\par\end{centering}

\caption{Signature of HOA and HOSPS in Binomial state for $M=20$.}

\end{figure}

\begin{figure}
\begin{centering}
\includegraphics{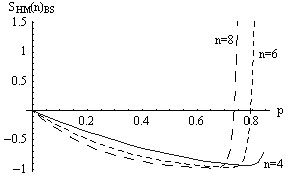}
\par\end{centering}

\caption{Signature of Hong Mandel Squeezing in Binomial State for $M=50.$}

\end{figure}

\subsection{Nonlinear Squeezed State}

Recently Darwish {[}\ref{the:Darwish}] has introduced a class of
nonlinear squeezed states. These states are named as nonlinear vacuum
squeezed state (NLVSS) and nonlinear first order excited squeezed
state (NLESS). Nonlinear states are expected to show nonclassical
properties. Keeping that in mind, Darwish has investigated the possibilities
of observing normal quadrature squeezing, amplitude squared squeezing
and antibunching. To be precise the study of Darwish shows that, NLESS
show subpoissonian photon statistics but does not show normal squeezing
and NLVSS does not show subpoissonian statistics but shows quadrature
squeezing. Here we wish to investigate the possibilities of observing
HOA and HOSPS in NLESS and NLVSS. Following Darwish we can define
\textcolor{black}{NLVSS} as \begin{equation}
|\psi\rangle_{V}=N\sum_{n'=0}^{\infty}\frac{\sqrt{(2n')!}}{n'![f(2n')]!}\left[\frac{\xi_{1}}{2}\right]^{n'}|2n'\rangle,\label{eq:ami4}\end{equation}
with \begin{equation}
|N|^{-2}=\sum_{n'=0}^{\infty}\frac{(2n')!}{\left(n'!\right)^{2}[f(2n')!]^{2}}\left[\frac{\xi_{1}}{2}\right]^{2n'}\label{eq:ami5}\end{equation}
and \textcolor{black}{NLESS} as \begin{equation}
|\phi\rangle_{E}=N^{\prime}\sum_{n'=0}^{\infty}\frac{\sqrt{(2n'+1)!}}{n![f(2n'+1)]!}\left[\frac{\xi_{1}}{2}\right]^{n'}|2n'+1\rangle,\label{eq:ami6}\end{equation}
 with \begin{equation}
|N^{\prime}|^{-2}=\sum_{n'=0}^{\infty}\frac{(2n'+1)!}{\left(n'!\right)^{2}[f(2n'+1)!]^{2}}\left[\frac{\xi_{1}}{2}\right]^{2n'}\label{eq:ami7}\end{equation}
where $f(.)$ is a well behaved nonunitary operator valued function
which is chosen in such a way that the normalization constant $N$
and $N^{\prime}$ must be bound. The above equations define NLESS
and NLVSS in general. To study the higher order nonclassical properties
of NLESS and NLVSS we have chosen a particular case in which $\xi_{1}=e^{i\phi}\tanh r$,
$\phi=\pi$ and $f(n')=\sqrt{n'}$. This particular case is considered
by Darwish {[}\ref{the:Darwish}] to study the possibility of observing
squeezing, amplitude squared squeezing and quasi probability distribution
nonlinear squeezed states. This particular choice of parameter yields
\begin{equation}
|N\,|=|N^{\prime}|=\left[\sum_{n'=0}^{\infty}\frac{1}{(n'!)^{2}}\left[\frac{\tanh r}{2}\right]^{2n'}\right]^{-\frac{1}{2}}.\label{eq:n=nprime}\end{equation}
 Using the definition of NLESS and the above choices of parameters
we obtain \begin{equation}
d(l)_{E}=|N\,|^{2}\sum_{n'=0}^{\infty}\frac{(2n'+1)!}{(2n'-l\,)!}\frac{1}{(n'!)^{2}}\left[\frac{\tanh r}{2}\right]^{2n'}-\left[|N\,|^{2}\sum_{n'=0}^{\infty}(2n'+1)\frac{1}{(n'!)^{2}}\left[\frac{\tanh r}{2}\right]^{2n'}\right]^{l+1},\label{eq:ami8a}\end{equation}

\begin{equation}
\begin{array}{lcl}
d_{h}(n-1)_{E} & = & \sum_{r=0}^{n}\sum_{k=1}^{r}\left[S_{2}(r,k)\,^{n}C_{r}(-1)^{r}\left(|N\,|^{2}\sum_{n'=0}^{\infty}(2n'+1)\frac{1}{(n'!)^{2}}\left[\frac{\tanh r}{2}\right]^{2n'}\right)^{n-r}\right.\\
 &  & \left.\left(|N\,|^{2}\sum_{n'=0}^{\infty}\frac{(2n'+1)!}{(2n'+1-k)!}\frac{1}{(n'!)^{2}}\left[\frac{\tanh r}{2}\right]^{2n'}-\left(|N\,|^{2}\sum_{n'=0}^{\infty}(2n'+1)\frac{1}{(n'!)^{2}}\left[\frac{\tanh r}{2}\right]^{2n'}\right)^{k}\right)\right]\end{array}\label{eq:ami8}\end{equation}

and\begin{equation}
\begin{array}{lcl}
S_{HM}(n)_{E} & = & \left[\frac{1}{\left(\frac{1}{2}\right)_{\frac{n}{2}}}\sum_{i=0}^{\frac{n}{2}}\sum_{k=0}^{n-2i}\frac{1}{2^{\frac{n}{2}}}t_{2i}\,^{n-2i}C_{k}\,^{n}C_{2i}\,|N\,|^{2}\right.\\
 &  & \left.\sum_{n'=Max\left[\frac{2k-n+2i}{2},\frac{n-2i-2k}{2}\right]}^{\infty}\frac{\sqrt{(2n'+1)!(2n'+1+n-2i-2k)!}}{(2n'+1-k\,)!\, n'!\left[\frac{2n'+n-2i-2k}{2}\right]!}\left[-\frac{\tanh r}{2}\right]^{{\normalcolor (4n'+n-2i-2k)/2}}\right]-1\end{array}\label{eq:ami9}\end{equation}
 Similarly we can write \begin{equation}
d(l)_{V}=|N\,|^{2}\sum_{n'=0}^{\infty}\frac{(2n')!}{(2n'-l-1)!}\frac{1}{(n'!)^{2}}\left[\frac{\tanh r}{2}\right]^{2n'}-\left[|N\,|^{2}\sum_{n'=0}^{\infty}(2n')\frac{1}{(n'!)^{2}}\left[\frac{\tanh r}{2}\right]^{2n'}\right]^{l+1},\label{eq:ami10a}\end{equation}
\begin{equation}
\begin{array}{lcl}
d_{h}(n-1)_{V} & = & \sum_{r=0}^{n}\sum_{k=1}^{r}\left[S_{2}(r,k)\,^{n}C_{r}(-1)^{r}\left(|N^{\prime}|^{2}\sum_{n'=0}^{\infty}(2n')\frac{1}{(n'!)^{2}}\left[\frac{\tanh r}{2}\right]^{2n'}\right)^{n-r}\right.\\
 &  & \left.\left(|N|^{2}\sum_{n'=0}^{\infty}\frac{(2n')!}{(2n'-k)!}\frac{1}{(n'!)^{2}}\left[\frac{\tanh r}{2}\right]^{2n'}-\left(|N|^{2}\sum_{n'=0}^{\infty}(2n')\frac{1}{(n'!)^{2}}\left[\frac{\tanh r}{2}\right]^{2n'}\right)^{k}\right)\right]\end{array}\label{eq:ami10b}\end{equation}
 and

\begin{equation}
\begin{array}{lcl}
S_{HM}(n)_{V} & = & \left[\frac{1}{\left(\frac{1}{2}\right)_{\frac{n}{2}}}\sum_{i=0}^{\frac{n}{2}}\sum_{k=0}^{n-2i}\frac{1}{2^{\frac{n}{2}}}t_{2i}\,^{n-2i}C_{k}\,^{n}C_{2i}\,|N\,|^{2}\right.\\
 &  & \left.\sum_{n'=Max\left[\frac{2k-n+2i}{2},\frac{n-2i-2k}{2}\right]}^{\infty}\frac{\sqrt{(2n')!(2n'+n-2i-2k)!}}{(2n'-k\,)!\, n'!\left[\frac{2n'+n-2i-2k}{2}\right]!}\left[-\frac{\tanh r}{2}\right]^{{\normalcolor (4n'+n-2i-2k)/2}}\right]-1\end{array}\label{eq:ami11}\end{equation}

Above expressions are graphically represented in Fig. 3-Fig. 7. From
Fig. 3 one can easily see that the condition for \textcolor{black}{fifth
order antibunching (HOA) and fifth order subpoissonian photon statistics
are satisfied by NLESS but the same is not satisfied by NLVSS (as
shown in Fig. 4). }Again Fig. 5 shows the absence of higher order
Hong Mandel squeezing in NLESS. However, Fig. 6 shows the presence
of higher order Hong Mandel squeezing in NLVSS. These observations
(Fig.3 -Fig 6) strongly establishes the fact that the HOS and HOSPS
are two independent phenomena. This observation is in accordance with
the corresponding lower order observations of Darwish {[}\ref{the:Darwish}]
related to these nonlinear squeezed states. Another interesting observation
is that the higher order squeezing parameter $(S_{HM}(n))$ oscillates
between nonclassical region and classical region in case of NLESS
for $n\geq10$. This oscillatory nature is depicted in Fig. 7

\begin{figure}
\begin{centering}
\includegraphics{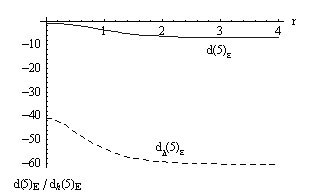}
\par\end{centering}

\caption{Signature of HOA and HOSPS in NLESS}

\end{figure}

\begin{figure}

\begin{centering}
\includegraphics{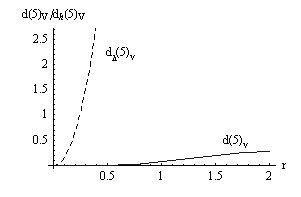}
\par\end{centering}

\caption{NLVSS shows superpoissonian characteristics}

\end{figure}

\begin{figure}

\begin{centering}
\includegraphics{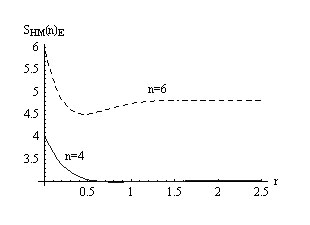}
\par\end{centering}

\caption{NLESS does not exhibit signature of HOS}

\end{figure}

\begin{figure}
\begin{centering}
\includegraphics{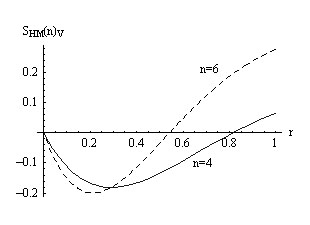}
\par\end{centering}

\caption{Signature of Hong Mandel squeezing in NLVSS}

\end{figure}

\begin{figure}
\begin{centering}
\includegraphics{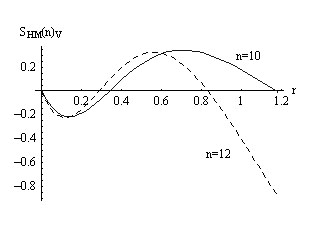}
\par\end{centering}

\caption{Oscillatory nature of Hong Mandel squeezing in NLVSS}

\end{figure}

\section{Conclusions}

\textcolor{black}{The criteria of HOSPS and Hong Mandel type of higher
order squeezing are derived from a single framework. Using that framework
and operator ordering theorem a} simpler form of the Hong-Mandel higher
order squeezing criterion is derived and generalized for \textcolor{black}{the}
multi-photon Bose operators of Brandt and Greenberg. \textcolor{black}{The
relation between HOA, HOSPS and HOS is investigated in detail and
certain interesting observations in this regard has been reported.
For example, it is shown that the lower order antibunching, HOA and
HOSPS appear in novel regimes (i.e. they may or may not appear simultaneously
as shown in Table 1). But in literature HOA and HOSPS have been used
as synonymous {[}\ref{the:Duc}].} Our observations establish that
it is incorrect to use the condition of HOA as a test of HOSPS. We
have used binomial state, \textcolor{black}{NLESS and NLVSS} as examples
of quantum state and have observed that BS always shows HOA and HOSPS
but it does not show HOS for all values of $p.$ So we conclude that
existence of HOSPS does not guarantee the existence of HOS. This is
consistent with the corresponding observations in lower order. Further,
it is also observed that the NLVSS which shows higher order squeezing
does not show HOSPS and HOA. The opposite is observed in NLESS and
consequently it is established that the HOSPS and HOS are two independent
signatures of higher order nonclassicality. Present study is the first
one of his kind in which rigorous attempts have been made to understand
the mutual relationship between different higher order nonclassical
states. The effort is successful to provide an insight into the mutual
relations between the well known nonclassical states and opens up
a possibility of similar work in broader class of nonclassical states.
The simpler framework provided for the study of possibilities of observing
Hong Mandel squeezing is also expected to be useful in the future
works.

\textbf{Acknowledgment}: AP thanks to DST, India for partial financial
support through the project grant SR\textbackslash{}FTP\textbackslash{}PS-13\textbackslash{}2004.
Authors are also thankful to Dr. B. P. Chamola and Mrs. Anindita Banerjee
for some useful discussions.

\end{document}